\documentclass[aps, prl, twocolumn, superscriptaddress]{revtex4-2}
\usepackage{bm, amsmath, amsfonts, amssymb, ascmac, mathtools}
\usepackage{multirow}
\usepackage{graphicx}
\usepackage{float, color, xcolor}
\usepackage{physics}
\usepackage{bbm}
\usepackage{tabularx}
\usepackage{enumerate}
\usepackage{comment}
\usepackage{braket}
\usepackage{bbm}

\usepackage[
pagebackref=false,
colorlinks=true,
linkcolor=blue,
urlcolor=blue,
filecolor=black,
citecolor=red,
pdfstartview=FitV,
pdftitle={},
pdfauthor={},
pdfsubject={},
pdfkeywords={},
pdfpagemode=None,
bookmarksopen=true
]{hyperref}

\newcommand{\ii}{\text{i}}

\begin{document}

\title{Koopman Nonlinear Non-Hermitian Skin Effect}

\author{Shu Hamanaka}
\email{hamanaka.shu.45p@st.kyoto-u.ac.jp}
\affiliation{Department of Physics, Kyoto University, Kyoto 606-8502, Japan}

\date{\today}

\begin{abstract}
Non-Hermitian skin effects are conventionally manifested as boundary localization of eigenstates in linear systems.
In nonlinear settings, however, where eigenstates are no longer well defined, it becomes unclear how skin effects should be faithfully characterized.
Here, we propose a Koopman-based characterization of nonlinear skin effects, in which localization is defined in terms of Koopman eigenfunctions in a lifted observable space, rather than physical states.
Using a minimal nonlinear extension of the Hatano-Nelson model, we show that dominant Koopman eigenfunctions localize sharply on higher-order observables, in stark contrast to linear skin effects confined to linear observables.
This lifted-space localization governs the sensitivity to boundary amplitude perturbations, providing a distinct dynamical signature of the nonlinear skin effect.
Our results establish the Koopman framework as a natural setting in which skin effects unique to nonlinear non-Hermitian systems can be identified.
\end{abstract}

\maketitle

Linearity plays a fundamental role in physics, as it enables a spectral decomposition of the Hamiltonian in terms of eigenvalues and eigenstates. This property allows complex time evolution to be reduced to a superposition of elementary modes, providing a powerful organizing principle across a wide range of physical systems~\cite{Sakurai}.

The non-Hermitian skin effect~\cite{Yao-Wang} is among the most prominent examples where a mode-based description plays a central role: nonreciprocity causes eigenstates to accumulate at system boundaries.
In linear non-Hermitian systems, the skin effect has been formulated in terms of non-Bloch band theory~\cite{Yokomizo-PRL-19,Amoeba} and point-gap topological characterizations~\cite{Zhang-PRL-20,Okuma-PRL-20}.
Here, the central quantities are again eigenvalues and eigenstates, or their generalizations such as pseudospectra~\cite{Okuma-PRB-20}.

Once nonlinearity is introduced, however, the notions of eigenstates and eigenvalues lose their direct meaning from the outset.
As a result, it becomes unclear how the skin effect should manifest itself in nonlinear systems, in particular whether boundary accumulation of stationary states alone provides an adequate diagnostic of the underlying nonlinear dynamics.
Previous approaches have mainly focused on stationary solutions, reducing the problem to a nonlinear spatial dynamical system at fixed frequency~\cite{Yuce-PLA-21, Yuce-PRB-25, Kawabata-PRL-25}.
While successful in certain contexts, such approaches may probe a restricted subset of the full nonlinear dynamics.
Consequently, the underlying nature of the nonlinear skin effect remains largely elusive.

\begin{figure}[t]
    \centering
    \includegraphics[width=1\linewidth]{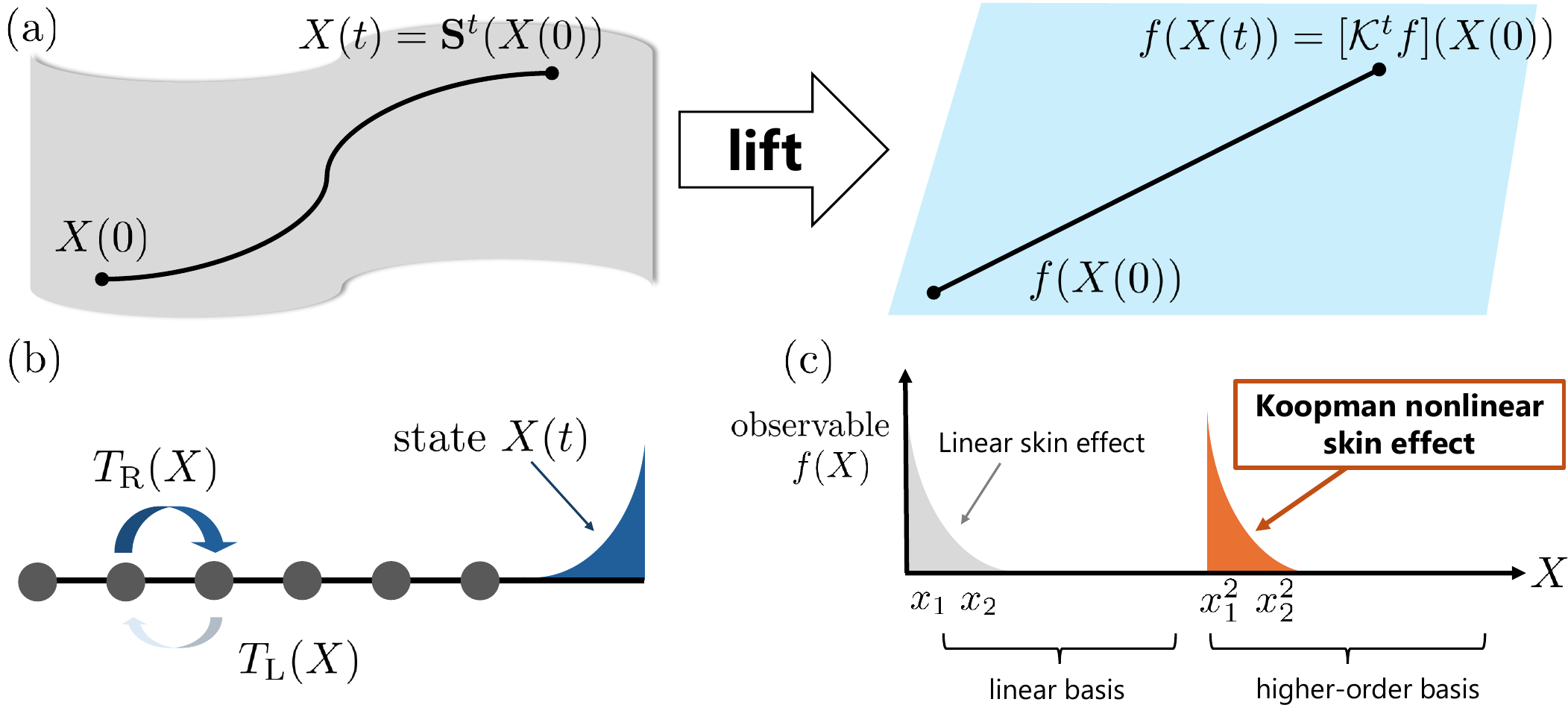}
    \caption{Illustration of the Koopman nonlinear non-Hermitian skin effect.
(a) The nonlinear Schr\"odinger dynamics is recast as a linear evolution in an appropriate observable space within the Koopman framework.
(b) The physical state $X(t)$ accumulates at the boundary due to nonreciprocal hopping, giving rise to the skin effect.
(c) Unlike the linear skin effect, where localization occurs on the linear observable sector, the Koopman nonlinear skin effect is characterized by the localization of observables on higher-order observable bases.
}
    \label{fig:model}
\end{figure}

In this Letter, we propose a Koopman-based characterization of nonlinear non-Hermitian skin effects, in which localization is diagnosed at the level of observables rather than physical states. 
Within this framework, we identify a nonlinear skin effect manifested as localization of dominant Koopman eigenfunctions in a lifted observable space (Fig.~\ref{fig:model}).
As an illustrative example, we study a minimal nonlinear extension of the Hatano-Nelson model and demonstrate that the dominant Koopman eigenfunctions localize sharply on higher-order observables in stark contrast to the linear case, where the dominant Koopman eigenfunctions are confined to linear observables.
We further introduce a dynamical probe based on the sensitivity to boundary amplitude perturbations, directly reflecting the localization property of Koopman eigenfunctions in the lifted observable space.
Our results establish a Koopman-based framework for identifying and probing skin effects unique to nonlinear non-Hermitian systems.

\textit{Koopman analysis}.---
Consider an $n$-component state vector $X=(x_1,\ldots,x_n)^{\mathsf T}\in\mathbb{C}^n$ evolving under a nonlinear equation
\begin{equation}\label{eq:nonlinear_schro}
\frac{d}{dt}X = \mathbf{T}(X),
\end{equation}
where $\mathbf{T}:\mathbb{C}^n\to\mathbb{C}^n$ is a generally nonlinear vector field.
Let $\mathbf{S}^t:\mathbb{C}^n\to\mathbb{C}^n$ denote the flow map generated by Eq.~\eqref{eq:nonlinear_schro}, defined by $\mathbf{S}^t(X(0))=X(t)$.  Equivalently, $\mathbf{S}^t$ satisfies $\partial_t \mathbf{S}^t(X)=\mathbf{T}(\mathbf{S}^t(X))$.

Koopman analysis shifts the focus from the state-space trajectory $X(t)$ to the evolution of \emph{observables}, i.e., scalar-valued functions $f:\mathbb{C}^n\to\mathbb{C}$ belonging to a chosen function space $\mathcal{F}$.
Given an observable $f\in\mathcal{F}$, its time evolution along the dynamics is the composition $f(X(t))=f(\mathbf{S}^t(X_0))$.  The Koopman operator $\mathcal{K}^t:\mathcal{F}\to\mathcal{F}$ is defined by~\cite{Koopman-31,Applied-Koopmanism,Mezic-13}
\begin{equation}\label{eq:koopman_def}
[\mathcal{K}^t f](X) = f(\mathbf{S}^t(X)),
\end{equation}
i.e., $\mathcal{K}^t$ maps an observable to another observable by composition with the flow. 
Crucially, although the underlying dynamics \eqref{eq:nonlinear_schro} is nonlinear, $\mathcal{K}^t$ acts \emph{linearly} on $\mathcal{F}$.

The spectral properties of $\mathcal{K}^t$ provide a linear representation of nonlinear dynamics.
Koopman eigenfunctions $\varphi_j\in\mathcal{F}$ satisfy
\begin{equation}
\mathcal{K}^t \varphi_j = e^{\lambda_j t}\varphi_j ,
\end{equation}
where $\lambda_j$ are the corresponding (continuous-time) Koopman eigenvalues.
Each eigenfunction $\varphi_j$ represents a dynamical mode whose temporal evolution is governed by $e^{\lambda_j t}$, allowing the long-time behavior of nonlinear dynamics to be organized in terms of dominant modes with the largest real parts of $\lambda_j$.

In the linear limit $\dot X = \mathbf{L}X$, a linear observable
$\varphi(X)=w^\dagger X$ with $w^\dagger$ being a left eigenvector of
$\mathbf{L}$ is a Koopman eigenfunction. Indeed,
\begin{align}
\mathcal{K}^t \varphi(X)
&= \varphi(e^{\mathbf{L}t}X)
= w^\dagger e^{\mathbf{L}t} X
= e^{\lambda t} \varphi(X),
\end{align}
where $w^\dagger \mathbf{L} = \lambda w^\dagger$.
This establishes a direct correspondence between the Koopman framework and the conventional spectral description in linear systems.

In practice, we simulate the continuous-time dynamics~\eqref{eq:nonlinear_schro} and sample the flow map as $X_{k+1}=\mathbf{S}^{\Delta t}(X_k)$ with a fixed time step $\Delta t$.
Applying extended dynamic mode decomposition (EDMD)~\cite{Williams-15,Korda-18,supplement} to snapshot pairs of observables then yields a finite-dimensional approximation of the Koopman operator $\mathcal{K}^{\Delta t}$ associated with this sampling time.

In this Letter, we exploit the structure of Koopman eigenfunctions in a lifted observable space to reveal a new type of non-Hermitian skin effect unique to nonlinear systems.

\textit{Model}.---
To identify a nonlinear skin effect within the Koopman framework, we study a nonlinear extension of the Hatano-Nelson model described by
\begin{equation}
    \ii \frac{d x_n}{dt}
    = T_{\rm R}(X)\, x_{n-1} + T_{\rm L}(X)\, x_{n+1},
    \label{eq:model}
\end{equation}
where $T_{\rm R}(X)$ and $T_{\rm L}(X)$ denote state-dependent nonreciprocal
hopping amplitudes
\begin{equation}
    T_{\rm R}(X) \!=\!t_{\rm R}\!\left[\!(1-\epsilon)\!+\!\epsilon\,\frac{x_{n-1}}{x_n}\right],\,
    T_{\rm L}(X)\!=\!t_{\rm L}\!\left[\!(1-\epsilon)\!+\!\epsilon\,\frac{x_{n+1}}{x_n}\right]
    \label{eq:nonlinear_hopping}
\end{equation}
with $\epsilon\in[0,1]$ controlling the strength of the nonlinearity.

In the linear limit $\epsilon=0$, the model reduces to the conventional
Hatano-Nelson Hamiltonian~\cite{Hatano-PRL-1996,Hatano-PRB-1997}.
For $t_{\rm R}>t_{\rm L}$, all eigenstates localize exponentially at the right boundary, giving rise to the linear skin effect, which has been extensively studied in a wide range of contexts~\cite{Bergholtz-RMP-18,Ashida-AdP-20,Okuma-review,Kunst-PRL-18,Lee-PRB-19,Zhang-PRL-20,Okuma-PRL-20,Okuma-PRL-21,Li-Ncome-20,Zhang-Ncom-22,Okugawa-PRB-20,Kawabata-PRB-20,Yoshida-PRR-20,Kawabata-PRL-21,Schomerus-PRR-20,Denner-Ncom-21,Haga-PRL-21,Mori-PRL-20,Schindler-PRB-21,Ma-PRR-24,Schindler-PRXQ-23,Nakamura-PRL-23,Hamanaka-PRL-24,Shen-Ncom-25,Nakagawa-PRX-25,Zhang-PRB-22,Kawabata-PRB-22,Faugno-PRL-22,Yoshida-PRL-24,Yuce-PLA-21,Yuce-PRB-25,Ezawa-PRB-22,Zhu-PRL-22,Many-PRB-24,Longhi-APR-25,Yoshida-PRB-25}.
This phenomenon has been experimentally observed in diverse platforms, including ultracold atoms~\cite{Liang-PRL-22,Wei-PRL-20,Zhao-Nature-25}, electric circuits~\cite{Helbig-Nphys-20,Hpfmann-PRR-20}, and photonic lattices~\cite{Weidemann-scieince-20}.
Beyond the linear regime, nonlinear extensions of the skin effect based on stationary nonlinear modes have also been demonstrated experimentally in photonic systems~\cite{Wang-PRL-25}.

In the nonlinear regime $\epsilon>0$, the hopping amplitudes acquire an explicit dependence on the instantaneous state $X$, rendering the dynamics nonlinear and spatially inhomogeneous.
Importantly, this nonlinearity does not introduce onsite terms but instead acts as a feedback mechanism that modifies the effective hopping processes in a configuration-dependent manner.

To analyze the resulting nonlinear dynamics, we perform a Koopman analysis by working in a lifted observable space.
Specifically, we consider a finite observable basis set $\{f_1,\ldots,f_{2L} \}=\{x_1,\ldots,x_L,x_1^2,\ldots,x_L^2\}$, which captures both linear and quadratic contributions of the state variables.
As shown below, this minimal lifting is sufficient in the present model to reveal dominant Koopman eigenfunctions exhibiting a nonlinear skin effect.

\begin{figure}[t]
    \centering
    \includegraphics[width=1\linewidth]{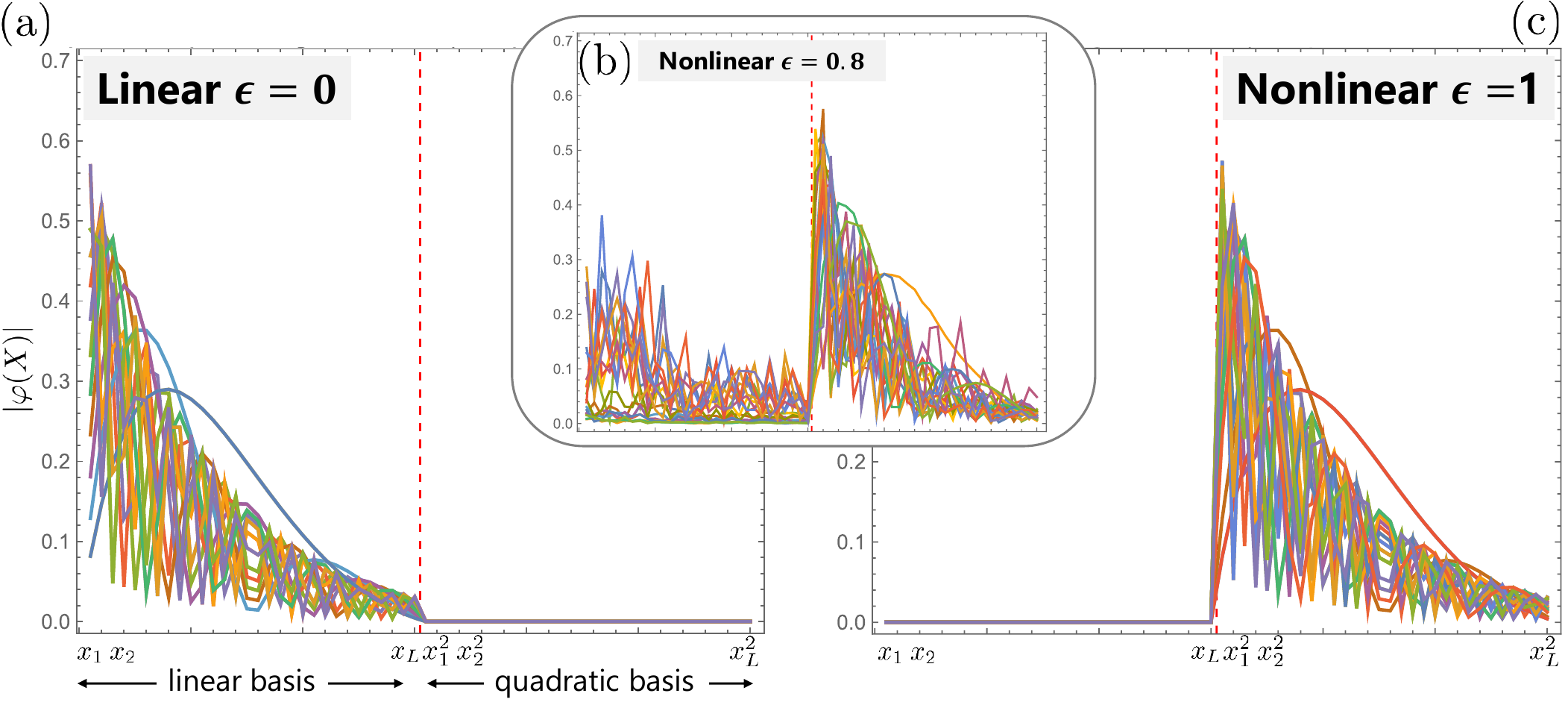}
    \caption{Koopman nonlinear non-Hermitian skin effect in the nonlinear Hatano-Nelson model ($L=30$, $ t_{\rm R}=0.55$, $t_{\rm L}=0.45$). (a) Linear regime ($\epsilon=0$): dominant Koopman eigenfunctions are localized on the linear observable $\{x_n\}$. (b) Intermediate nonlinearity ($\epsilon=0.8$): the dominant Koopman eigenfunctions acquire a strong weight on the quadratic observables $\{x_n^2\}$, with only subleading contributions from the linear sector. (c) Exactly solvable nonlinear limit ($\epsilon=1$): the dominant Koopman eigenfunctions are fully localized on the higher-order (quadratic) observable $\{x_n^2\}$, demonstrating that the essential dynamical degrees of freedom are shifted to higher-order observables. In all panels, we show the twenty slowest-decaying modes, obtained using EDMD~\cite{supplement}.
}
    \label{fig:linear nonlinear}
\end{figure}

\textit{Linear skin effect}.---
We first consider the linear limit $\epsilon=0$, in which the linear generator coincides with that of the conventional Hatano--Nelson model.
In this case, Koopman eigenfunctions associated with linear observables $f_n(X)=x_n$ coincide with the left eigenstates of the Hatano-Nelson Hamiltonian. For $t_{\rm R}>t_{\rm L}$, these left eigenstates are
exponentially localized toward the left boundary, and so are the dominant
Koopman eigenfunctions [Fig.~\ref{fig:linear nonlinear} (a)]. This provides a
Koopman representation of the conventional linear non-Hermitian skin effect.

Importantly, this localization should not be confused with the boundary
accumulation of the physical state itself: while right eigenstates of the
Hatano-Nelson Hamiltonian localize toward the right boundary, Koopman
eigenfunctions live in the observable space and inherit their localization from the left eigenstates of the linear generator.
This localization implies that the dominant Koopman eigenfunctions place a large weight on observables associated with the left boundary, indicating that the long-time dynamics is organized by modes that are particularly sensitive to the initial state near the left edge.

\textit{Koopman nonlinear non-Hermitian skin effect}.---
We now turn to the nonlinear regime $\epsilon>0$.
For intermediate nonlinearity $0<\epsilon<1$, we find that the dominant Koopman eigenfunctions are no longer localized on the linear observables $\{x_n\}$, but instead acquire a strong weight on quadratic observables $\{x_n^2\}$.
Figure~\ref{fig:linear nonlinear} (b) shows a representative result at $\epsilon=0.8$, where the Koopman eigenfunctions are predominantly localized in the lifted quadratic sector, with only subleading contributions from the linear observables.
This demonstrates that the essential dynamical degrees of freedom are shifted from linear observables to higher-order observables by the nonlinearity.

The origin of this behavior becomes transparent in the exactly solvable
limit $\epsilon=1$.
In this limit, the nonlinear Schr\"odinger equation~\eqref{eq:model}
induces a closed linear evolution for the quadratic observables,
\begin{equation}
\ii \frac{d f_n}{dt}
= 2 t_{\rm R} f_{n-1} + 2 t_{\rm L} f_{n+1},
\qquad (f_n \equiv x_n^2),
\label{eq:epsilon=1}
\end{equation}
which defines a linear Hatano-Nelson dynamics in the lifted observable
space.
The corresponding Koopman generator therefore admits eigenfunctions of
the form
\begin{equation}
\varphi_j(X)=\sum_n w^{(j)}_n x_n^2,
\label{eq:koopman_phi}
\end{equation}
with $w^{(j)}_n \propto (\sqrt{t_{\rm L}/t_{\rm R}})^n \sin(n\theta_j)$,
$\theta_j=j\pi/(L+1)$, and Koopman eigenvalue
$\lambda_j = -4 \ii \sqrt{t_{\rm R}t_{\rm L}}\cos\theta_j$.
As a result, the Koopman eigenfunctions exhibit exponential boundary
localization in the lifted observable space, constituting a
Koopman-based nonlinear skin effect [Fig.~\ref{fig:linear nonlinear} (c)].

Taken together, from a unified Koopman perspective, linear and nonlinear skin effects differ in the observable sector where localization occurs: the linear skin effect is confined to linear observables, whereas the Koopman nonlinear skin effect manifests itself as localization in higher-order observables.

\begin{figure}[t]
    \centering
    \includegraphics[width=1\linewidth]{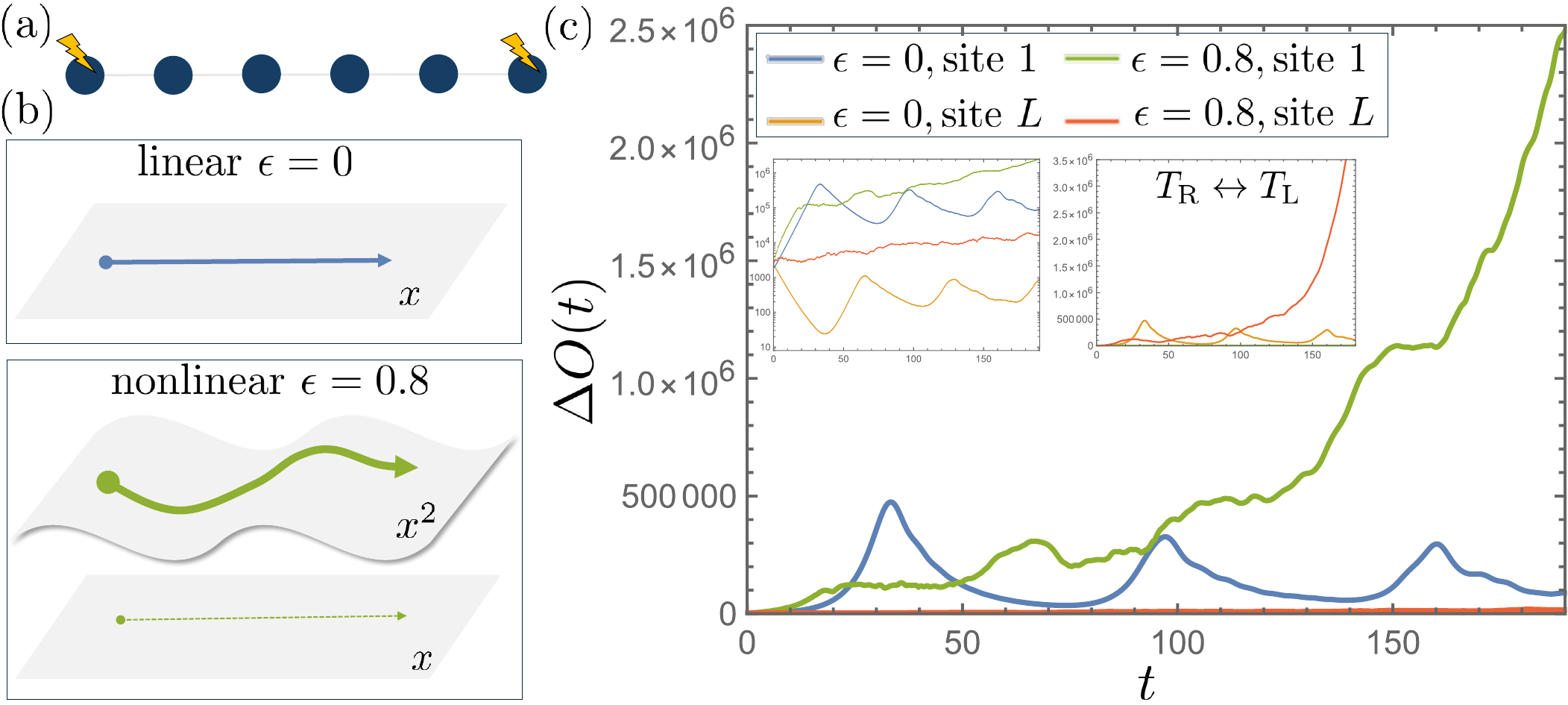}
    \caption{Dynamical manifestation of the Koopman nonlinear skin effect ($L=30$, $t_{\rm R}=0.55$, $t_{\rm L}=0.45$).
(a) Schematic illustration of the boundary perturbation protocol.
(b) Conceptual comparison between linear ($\epsilon=0$) and nonlinear ($\epsilon=0.8$) dynamics.
(c) Time evolution of $\Delta O(t)$ defined in Eq.~\eqref{eq:deltaO2}. Blue and orange curves correspond to perturbations applied at the left
($j=1$) and right ($j=L$) boundaries, respectively, in the linear regime
($\epsilon=0$), while green and red curves show the corresponding results for $\epsilon=0.8$. The left inset shows the same data plotted on a logarithmic scale. The right inset shows the result obtained after reversing the direction of nonreciprocity ($T_{\rm R}\leftrightarrow T_{\rm L}$), ruling out a generic dynamical instability.
}
\label{fig:sensitivity}
\end{figure}

\textit{Boundary-amplitude sensitivity as a probe of the Koopman skin effect}.---
We now demonstrate how the Koopman skin effect identified above, namely the localization of dominant Koopman eigenfunctions on the higher-order observable basis, manifests itself dynamically and can be probed through sensitivity to boundary amplitude perturbations.

As shown in Fig.~\ref{fig:linear nonlinear}, Koopman eigenfunctions are confined to the linear observable sector in the linear regime ($\epsilon=0$), whereas for $\epsilon>0$ they acquire dominant weight on the quadratic sector with pronounced boundary localization.
This localization property suggests that perturbations acting primarily on quadratic observables can selectively probe the nonlinear Koopman skin modes.

To substantiate this idea, we employ the global quadratic quantity $O(t) \equiv \sum_{n=1}^{L} x_n^2(t)$, which is spatially uniform and does not rely on any a priori knowledge of Koopman eigenfunctions.
We prepare two nearby initial conditions that differ only at a boundary site, $x_j(0)=A$ and $x_j(0)=A+\delta$, with $A=10^3$ and $\delta=1$ [Fig.~\ref{fig:sensitivity} (a)], and monitor their separation through
\begin{equation}\label{eq:deltaO2}
\Delta O(t) \equiv \bigl| O^{A+\delta}(t) - O^{A}(t) \bigr|.
\end{equation}

Figure~\ref{fig:sensitivity} (c) shows the time evolution of $\Delta O(t)$ for perturbations applied at the left and right boundaries, for both $\epsilon=0$ and $\epsilon=0.8$.
In the linear regime, perturbations at either boundary lead only to weak transient responses without sustained growth [Fig.~\ref{fig:sensitivity} (c), blue and orange curves].
This reflects the fact that the dynamics is faithfully organized by linear observables [Fig.~\ref{fig:sensitivity} (b)], so that the initial difference $(A+\delta)-A=\delta \ll A$ remains negligible throughout the time evolution.

In contrast, in the nonlinear regime, a perturbation applied at the left boundary produces a pronounced and sustained growth of $\Delta O(t)$ [Fig.~\ref{fig:sensitivity} (c), green curve].
At the quadratic level, the initial difference scales as $(A+\delta)^2-A^2 \simeq A\delta$, so that even a small boundary perturbation is strongly amplified in the lifted observable space [Fig.~\ref{fig:sensitivity} (b)].
Crucially, this amplification is boundary selective: a perturbation applied at the right boundary results in only a negligible response [Fig.~\ref{fig:sensitivity} (c), red curve].
This asymmetry directly reflects the boundary localization of the dominant Koopman eigenfunctions, which carry negligible weight near the right boundary.

Taken together, these results demonstrate that sensitivity to boundary amplitude perturbations is governed by the localization properties of Koopman eigenfunctions in a lifted observable space.
A Koopman-based viewpoint, rather than conventional stationary-state analysis, is therefore essential for understanding the dynamical behavior observed here.

\textit{Generality of the Koopman nonlinear skin effect}.---
So far, we have focused on the simplest model in order to demonstrate the essence of the Koopman nonlinear skin effect in a transparent and analytically controlled manner. The phenomenon identified here is not restricted to such minimal or exactly solvable models, but arises more generally in nonlinear non-Hermitian systems.

To illustrate this, we consider another prototypical model originally introduced in Ref.~\cite{Kawabata-PRL-25}
\begin{align}\label{eq:KK-model}
    H = \sum_n \Big( -&\frac{1-\gamma + \epsilon \abs{x_n}^2}{2} \ket{n}\bra{n+1} \nonumber \\
    &\quad -\frac{1+\gamma - \epsilon \abs{x_{n+1}}^2}{2} \ket{n+1}\bra{n} \Big),
\end{align}
where nonreciprocal hopping is combined with nonlinear feedback through the local amplitude, rendering the dynamics nonlinear and configuration dependent.

\begin{figure}[t]
    \centering
    \includegraphics[width=1\linewidth]{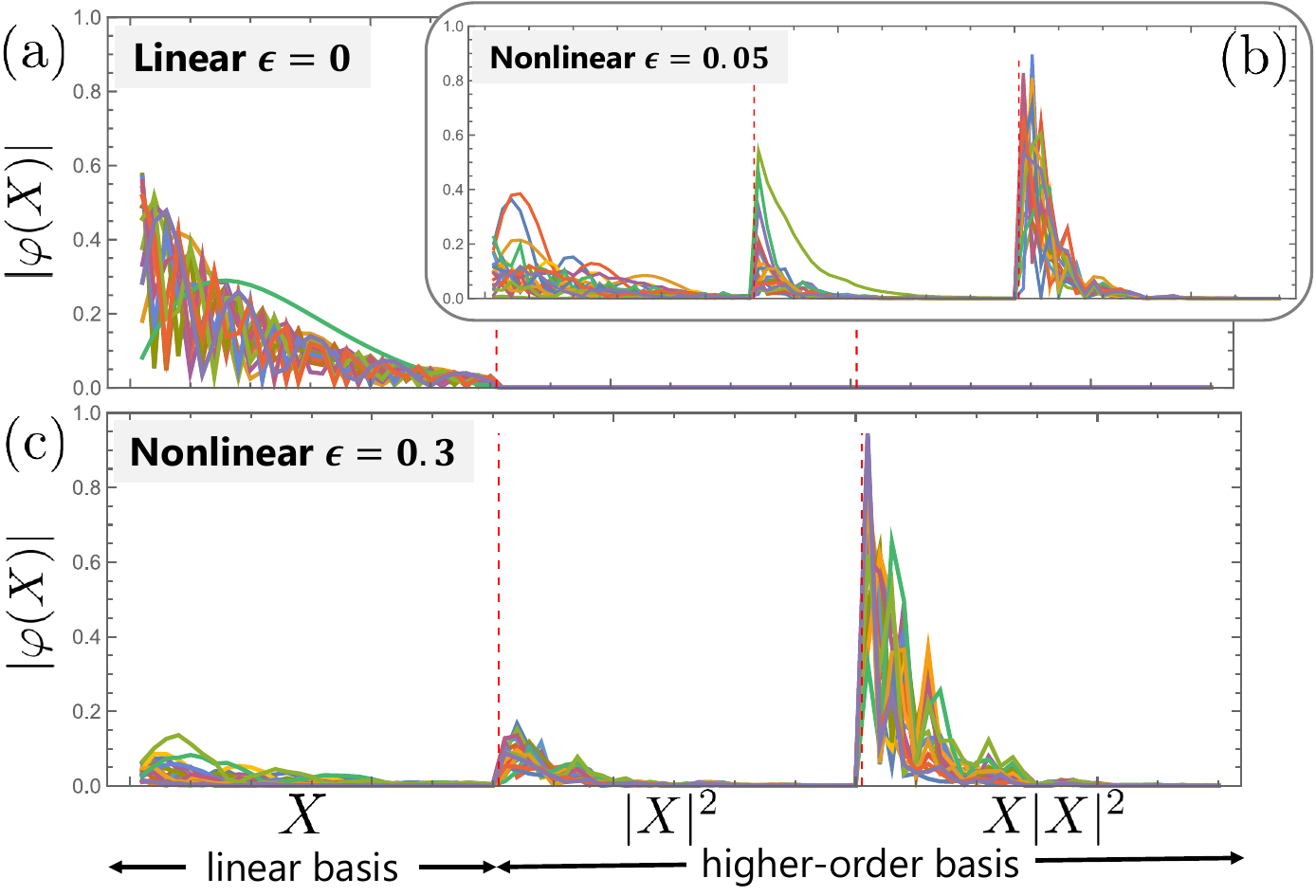}
    \caption{Generality of the Koopman nonlinear skin effect ($L=30,\gamma=0.1$). Distribution of the dominant Koopman eigenfunctions in the lifted observable space for the model defined in Eq.~\eqref{eq:KK-model}.
(a) Linear regime ($\epsilon = 0$): the Koopman eigenfunctions are localized on the linear observables. (b) Weak nonlinearity ($\epsilon = 0.05$): the Koopman eigenfunctions acquire finite weight on higher-order observables. (c) Stronger nonlinearity ($\epsilon = 0.3$): the dominant Koopman eigenfunctions are predominantly localized on higher-order observables. In all panels, we show the twenty slowest-decaying modes, obtained using EDMD~\cite{supplement}.}
    \label{fig:KKmodel}
\end{figure}

In the linear limit $\epsilon = 0$, the dynamics is fully captured by the linear observables $\{x_n\}$, and the dominant Koopman eigenfunctions are localized on linear bases, reflecting the linear non-Hermitian skin effect [Fig.~\ref{fig:KKmodel} (a)].

Once nonlinearity is introduced ($\epsilon>0$), the effective degrees of freedom governing the long-time dynamics progressively shift from linear to higher-order observables.
Correspondingly, the dominant Koopman eigenfunctions acquire substantial weight in the higher-order observable sector, and their support moves toward increasingly higher-order bases as the nonlinearity is strengthened, inducing the Koopman nonlinear skin effect [Fig.~\ref{fig:KKmodel} (b, c)].
This redistribution of dynamical weight within the observable space closely parallels that observed in the minimal model studied above, even though the lifted observable basis does not form an exactly closed subspace under the nonlinear dynamics.

Therefore, these results indicate that the Koopman nonlinear skin effect is a generic phenomenon in nonlinear non-Hermitian systems, arising from the redistribution of dynamical weight in observable space induced by nonlinearity.

\textit{Discussion}.---
In this Letter, we establish a Koopman-based characterization of nonlinear non-Hermitian skin effects, in which localization is defined at the level of observables, rather than physical states.
Within this framework, the skin effect manifests itself as localization of Koopman eigenfunctions in a lifted observable space.
Using a minimal nonlinear model, we show that the dominant modes localize on higher-order observable bases, and that this lifted-space localization directly governs the sensitivity to boundary amplitude perturbations.

Our results suggest that, while stationary-state analysis captures certain aspects of nonlinear skin effects, there exists an underlying dynamical structure that is naturally revealed in the lifted observable space.
Accordingly, the structure of the lifted observable space plays a central role in organizing the nonlinear dynamics, providing a natural setting in which skin effects unique to nonlinear non-Hermitian systems can be identified.

While the present work focuses on Koopman eigenfunctions, it is worth noting that, in linear non-Hermitian systems, the skin effect is closely tied to point-gap topology~\cite{Zhang-PRL-20,Okuma-PRL-20}.
In the linear regime, such topological characterizations can be naturally reformulated within the Koopman framework, since the Koopman spectrum is directly linked to the eigenvalue of the Hamiltonian.
Whether and how point-gap topology admits a meaningful generalization in genuinely nonlinear dynamics, however, remains an open question, which we leave for future work.

\medskip
\begingroup
\renewcommand{\addcontentsline}[3]{}
\begin{acknowledgments}
This work was supported by JSPS Research Fellow No.~24KJ1445.
\end{acknowledgments}
\endgroup

\let\oldaddcontentsline\addcontentsline
\renewcommand{\addcontentsline}[3]{}
\bibliography{ref.bib}
\let\addcontentsline\oldaddcontentsline

\clearpage
\widetext

\setcounter{secnumdepth}{3}

\renewcommand{\theequation}{S\arabic{equation}}
\renewcommand{\thefigure}{S\arabic{figure}}
\renewcommand{\thetable}{S\arabic{table}}
\setcounter{equation}{0}
\setcounter{figure}{0}
\setcounter{table}{0}
\setcounter{section}{0}
\setcounter{tocdepth}{0}

\numberwithin{equation}{section} 

\begin{center}
{\bf \large Supplemental Material for \\ \smallskip 
``Koopman Nonlinear Non-Hermitian Skin Effect"}
\end{center}



\section{Extended Dynamic Mode Decomposition (EDMD)}

\subsection{General framework}
\label{sec:SM_EDMD_general}
We briefly review the extended dynamic mode decomposition (EDMD)~\cite{Williams-15,Korda-18}, a standard data-driven method to approximate the Koopman operator in a finite-dimensional lifted observable space.
Consider a nonlinear dynamical system
\begin{equation}
    \frac{dX}{dt} = \mathbf{T}(X), \qquad X \in \mathbb{C}^n,
    \label{eq:SM_general_dynamics}
\end{equation}
with a generally nonlinear vector field $\mathbf{T}(X)$.
Let $\mathbf{S}^t$ denote the flow map generated by Eq.~\eqref{eq:SM_general_dynamics},
defined by $X(t)=\mathbf{S}^t(X(0))$.
Sampling the continuous-time dynamics with a fixed time interval $\Delta t$
yields a discrete-time map
\begin{equation}
    X_{k+1} = \mathbf{S}^{\Delta t}(X_k).
\end{equation}

The Koopman operator $K^{\Delta t}$ acts linearly on observables
$g:\mathbb{C}^n\to\mathbb{C}$ as
\begin{equation}
    [K^{\Delta t} g](X) = g(\mathbf{S}^{\Delta t}(X)).
\end{equation}
EDMD approximates $K^{\Delta t}$ in a finite-dimensional observable space
spanned by a chosen dictionary
$\{g_1,\ldots,g_p\}$.

Let $\bm g(X)=(g_1(X),\ldots,g_p(X))^{\mathsf T}\in\mathbb{C}^p$.
Given snapshot pairs $\{(Y_m,Z_m)\}_{m=1}^M$ with
$Z_m=\mathbf{S}^{\Delta t}(Y_m)$, we construct the data matrices
\begin{equation}
    Y_{m,:}=\bm g(Y_m)^{\mathsf T},\qquad
    Z_{m,:}=\bm g(Z_m)^{\mathsf T}.
\end{equation}
The EDMD approximation of the Koopman operator is defined as the solution of
the least-squares problem
\begin{equation}
    K = \arg\min_{K'} \|Z-YK'\|_F^2,
\end{equation}
which yields
\begin{equation}
    K=(Y^\dagger Y+\eta_{\rm reg}\mathbb{I})^{-1}Y^\dagger Z
\end{equation}
with Tikhonov regularization $\eta_{\rm reg}>0$.
Diagonalization of $K$ yields the discrete-time Koopman eigenvalues
$\Lambda_j$ and the corresponding left eigenvectors $w_j^\dagger$,
from which the Koopman eigenfunctions are approximated as
\begin{equation}\label{aeq:varphi}
    \varphi_j(X)\approx w_j^\dagger \bm g(X).
\end{equation}
We note that the left eigenvectors $w_j^\dagger$ appearing in Eq.~\eqref{aeq:varphi} are those of the finite-dimensional EDMD matrix $K$ acting on the chosen observable dictionary.
In the linear limit, where the observable dictionary consists only of linear observables, the EDMD construction reduces to the conventional linear Koopman description.
Accordingly, the EDMD eigenvectors $w_j^\dagger$ coincide with the left eigenvectors of the linear generator discussed in the main text, ensuring consistency between the EDMD-based analysis and the linear theory.
Note that the corresponding continuous-time growth rates are obtained via
$\lambda_j=\Delta t^{-1}\log\Lambda_j$.

In the main text, we applied the above framework to two nonlinear
non-Hermitian models.
In the following, we present supplementary details of the EDMD implementation for each case.

\subsection{Application to the minimal model}
\label{sec:SM_EDMD_minimal}
We first describe the details of the EDMD implementation for the minimal nonlinear Hatano-Nelson model defined in Eqs.~\eqref{eq:model} and~\eqref{eq:nonlinear_hopping} of the main text.
The flow map $\mathbf{S}^{\Delta t}$ is generated by integrating the differential equation over the interval $[0,\Delta t]$.

\paragraph{Parameters and sampling.}
We use a chain of length $L=30$ and sample the dynamics with
$\Delta t=0.2$.
A trajectory of length $N_{\rm steps}=1000$ is generated, and a short
transient of $N_{\rm trans}=1$ step is discarded.
The EDMD regularization parameter is fixed to
$\eta_{\rm reg}=10^{-10}$.

\paragraph{Observable dictionary.}
We employ the lifted observable map
\begin{equation}
    \bm g(X)=(x_1,\ldots,x_L,x_1^2,\ldots,x_L^2)^{\mathsf T}.
\end{equation}
In the exactly solvable limit $\epsilon=1$, this observable subspace is
closed.
For $0<\epsilon<1$, the same dictionary provides an effective reduced
representation of the dominant dynamics.

\paragraph{Initial condition.}
We initialize a random state with controlled amplitude,
\begin{equation}
    x_j(0)\sim \text{Uniform }[0.1,0.2],\qquad
    x_{\frac{L}{2}}(0)\mapsto x_{\frac{L}{2}}(0)+1,
\end{equation}
followed by normalization \(X(0)\mapsto X(0)/\|X(0)\|\).
This choice ensures that the initial state has a finite amplitude over the entire chain to avoid the divergence while introducing a localized perturbation at the center site, which is convenient for probing the subsequent redistribution of dynamical
weight in the lifted observable space.

\subsection{Application to the model in Ref.~\cite{Kawabata-PRL-25}}
\label{sec:SM_EDMD_KK}
We next consider the nonlinear Hatano-Nelson model in Eq.~\eqref{eq:KK-model}, which was originally introduced in Ref.~\cite{Kawabata-PRL-25}.

\paragraph{Parameters and sampling.}
We consider a chain of length $L=30$ with nonreciprocity parameter
$\gamma=0.1$.
The flow map is sampled with $\Delta t=0.1$, generating a trajectory of
length $N_{\rm steps}=1500$ after discarding a transient of
$N_{\rm trans}=1$ step.
The regularization parameter is chosen as
$\eta_{\rm reg}=10^{-10}$.

\paragraph{Observable dictionary.}
For this model, we use an extended dictionary,
\begin{equation}
    \bm g(X)
    =
    (x_1,\ldots,x_L,
    |x_1|^2,\ldots,|x_L|^2,
    |x_1|^2x_1,\ldots,|x_L|^2x_L)^{\mathsf T},
\end{equation}
which is found to provide a stable and physically transparent description
of the nonlinear dynamics.
Unlike the minimal model, this observable space is not closed under the
exact Koopman evolution.

\paragraph{Initial condition.}
For this model, we prepare the initial state as a random vector with controlled amplitude,
\begin{equation}
    x_j(0)\sim \text{Uniform }[0.1,0.2],\qquad
    x_{\frac{L}{2}}(0)\mapsto x_{\frac{L}{2}}(0)+3,
\end{equation}
followed by normalization
$X(0)\mapsto X(0)/\|X(0)\|$.

\section{Validation of the EDMD approximation}
\label{sec:SM_EDMD_validation}
\begin{figure}[t]
    \centering
    \includegraphics[width=1\linewidth]{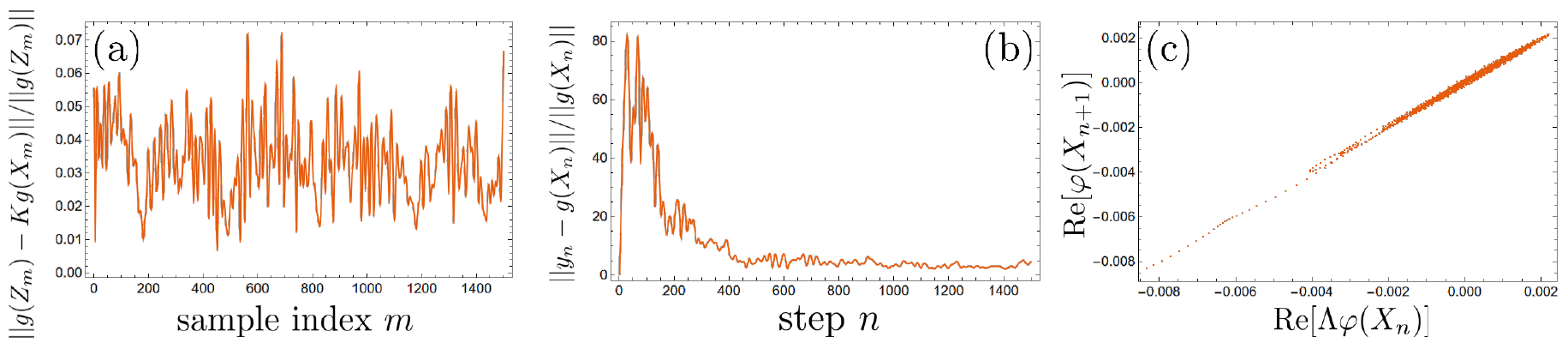}
    \caption{Validation of the EDMD approximation ($L=30, \gamma=0.1,\epsilon=0.05$).
(a) Per-sample relative one-step residual $\|\bm g(Z_m)-K\bm g(Y_m)\|_2/\|\bm g(Z_m)\|_2$ on the training data.
(b) Multi-step rollout error $E_{\mathrm{roll}}(n)=\|y_n-\bm g(X_n)\|_2/\|\bm g(X_n)\|_2$ for an unseen initial condition.
(c) Eigenfunction consistency check for the dominant Koopman mode: scatter plot of $\Re[\varphi(X_{n+1})]$ versus $\Re[\Lambda\varphi(X_n)]$.}
\label{fig:kk_edmd_validation}
\end{figure}

For the second model analyzed in the main text, originally introduced in Ref.~\cite{Kawabata-PRL-25}, the observable basis is not closed under the Koopman dynamics for any $\epsilon>0$, in contrast to the minimal model.
As a result, an exact benchmark for the EDMD approximation of the Koopman operator is not available.
In this section, we therefore quantitatively validate the accuracy of the EDMD approximation.
Figure~\ref{fig:kk_edmd_validation} summarizes three complementary diagnostics, which we explain below.

\paragraph{(a) One-step operator residual.}
From the sampled trajectory $\{Y_m,Z_m\}$, with $Z_m=\mathbf{S}^{\Delta t}(Y_m)$,
EDMD determines the Koopman matrix $K$ via the least-squares problem $Z \simeq YK$, where
$Y_{m,:}=\bm g(Y_m)^{\mathsf T}$ and $Z_{m,:}=\bm g(Z_m)^{\mathsf T}$.
To assess the quality of this approximation, we evaluate the per-sample relative one-step residual
\begin{equation}
e^{(1)}_m
=
\frac{\|\bm g(Z_m)-K\,\bm g(Y_m)\|_2}{\|\bm g(Z_m)\|_2}.
\end{equation}

Figure~\ref{fig:kk_edmd_validation} (a) shows the per-sample one-step
relative residual $e^{(1)}_m$.
The error remains bounded over all snapshots and does not exceed
$e^{(1)}_m \approx 0.07$.
Such a uniformly small one-step residual demonstrates that the EDMD
matrix $K$ accurately captures the local Koopman evolution on the
training data, even though the chosen observable basis is not closed
under the nonlinear dynamics.

\paragraph{(b) Multi-step rollout from an unseen initial condition.}
To test generalization beyond the training trajectory, we consider an
independent initial condition $X_0^{\mathrm{test}}$ not used in the EDMD
construction.
Starting from $y_0=\bm g(X_0^{\mathrm{test}})$, we propagate the lifted state via the linear recurrence $y_{n+1}=K y_n$ using the EDMD-approximated Koopman matrix $K$, and compare it with the true lifted trajectory $\bm g(X_n)$ generated by the full nonlinear dynamics.
We evaluate the relative error from
\begin{equation}
E_{\mathrm{roll}}(n)
=
\frac{\|y_n-\bm g(X_n)\|_2}{\|\bm g(X_n)\|_2}.
\end{equation}
Figure~\ref{fig:kk_edmd_validation} (b) shows the multi-step rollout error
for an initial condition not used in the EDMD construction.
After a transient increase at early times, the error remains bounded and decreases to a small value at longer times, reflecting the contractive nature of the dominant Koopman modes in this parameter regime.
This behavior confirms that the EDMD operator provides a faithful local linear approximation of the nonlinear dynamics.

\paragraph{(c) Koopman eigenfunction consistency.}
Finally, we directly test the defining relation of Koopman eigenfunctions.
From a left eigenvector $w^\dagger$ of $K$ with eigenvalue $\Lambda$,
we construct the corresponding Koopman eigenfunction
$\varphi(X)=w^\dagger \bm g(X)$.
For an exact Koopman eigenfunction, one should have
$\varphi(X_{n+1})=\Lambda\varphi(X_n)$.
Figure~\ref{fig:kk_edmd_validation} (c) shows a scatter plot of
$\Re[\varphi(X_{n+1})]$ versus $\Re[\Lambda\varphi(X_n)]$ for the dominant mode.
The strong linear correlation demonstrates that this mode satisfies the
Koopman eigenfunction relation to a good approximation, validating its
interpretation as a genuine Koopman mode of the nonlinear dynamics.

Taken together, these diagnostics establish that EDMD yields a reliable
finite-dimensional approximation of the Koopman operator for this model, providing a solid foundation for the analysis of the Koopman nonlinear non-Hermitian skin effect presented in the main text.

\end{document}